\RequirePackage{lineno}

\documentclass[twocolumn,aps,prc,showpacs,superscriptaddress,floatfix]{revtex4}

\usepackage{color}
\usepackage{graphicx}
\usepackage{multirow}

\newcommand {\gevc}	{GeV/$c$}

\newcommand {\fmc}	{fm/$c$}
\newcommand {\Ru}	{$^{96}_{44}$Ru}
\newcommand {\Zr}	{$^{96}_{40}$Zr}

\newcommand {\Nch}	{N_{\rm ch}}
\newcommand {\dNdeta}	{d\Nch/d\eta}
\newcommand {\Npart}	{N_{\rm part}}
\newcommand {\Nbin}	{N_{\rm bin}}
\newcommand {\minv}	{m_{\rm inv}}
\newcommand {\mcg}	{MC-Glb}
\newcommand {\hij}	{HIJING}
\newcommand {\ampt}	{AMPT}
\newcommand {\amptsm}	{AMPT-SM}
\newcommand {\wsQ}	{WS-R$_{\rm Q}$}
\newcommand {\wsR}	{WS-R$_{\rm DFT}$}
\newcommand {\amptwsQ}	{\amptsm(\wsQ)}
\newcommand {\amptwsR}	{\amptsm(\wsR)}
\newcommand {\amptdft}	{\amptsm(DFT)}

\newcommand {\rhows}	{\rho_{_{\rm WS}}}
\newcommand {\etwo}	{\epsilon_2}

\newcommand {\rt}	{r_{T}}
\newcommand {\pt}	{p_{T}}

\newcommand {\mean}[1]	{\langle #1\rangle}

\begin{document}
\title{Multiphase transport model predictions of isobaric collisions with nuclear structure from density functional theory}
\author{Hanlin Li}
\email{lihl@wust.edu.cn}
\affiliation{College of Science, Wuhan University of Science and Technology, Wuhan, Hubei 430065, China}
\author{Hao-jie Xu}
\email{haojiexu@zjhu.edu.cn}
\affiliation{School of Science, Huzhou University, Huzhou, Zhejiang 313000, China}
\author{Jie Zhao}
\email{zhao656@purdue.edu}
\affiliation{Department of Physics and Astronomy, Purdue University, West Lafayette, Indiana 47907, USA}
\author{Zi-Wei Lin}
\affiliation{Key Laboratory of Quarks and Lepton Physics (MOE) and Institute of Particle Physics, Central China Normal University, Wuhan, Hubei 430079, China}
\affiliation{Department of Physics, East Carolina University, Greenville, North Carolina 27858, USA}
\author{Hanzhong Zhang}
\affiliation{Key Laboratory of Quarks and Lepton Physics (MOE) and Institute of Particle Physics, Central China Normal University, Wuhan, Hubei 430079, China}
\author{Xiaobao Wang}
\affiliation{School of Science, Huzhou University, Huzhou, Zhejiang 313000, China}
\author{Caiwan Shen}
\affiliation{School of Science, Huzhou University, Huzhou, Zhejiang 313000, China}
\author{Fuqiang Wang}
\email{fqwang@zjhu.edu.cn}
\affiliation{School of Science, Huzhou University, Huzhou, Zhejiang 313000, China}
\affiliation{Department of Physics and Astronomy, Purdue University, West Lafayette, Indiana 47907, USA}

\date{\today}

\begin{abstract}
  Isobaric \Ru+\Ru\ and \Zr+\Zr\ collisions were performed at the Relativistic Heavy Ion Collider in 2018. Using the "a multi-phase transport" model with nuclear structures calculated by the density functional theory (DFT), we make predictions for the charged hadron multiplicity distributions
  and elliptic azimuthal anisotropies in these collisions. Emphases are put on the relative differences between the two collision systems that can decisively discriminate DFT nuclear distributions from the commonly used Woods-Saxon densities.
\end{abstract}


\maketitle

\section{Introduction}
The isobar run in 2018, colliding \Ru\ (Ru+Ru) and \Zr\ (Zr+Zr) nuclei at the Relativistic Heavy-Ion Collider (RHIC), was motivated by the search for the chiral magnetic effect (CME) in quantum chromodynamics~\cite{Kharzeev:2015znc,Skokov:2016yrj,Zhao:2018ixy}. Because of the different numbers of protons, the CME is expected to differ between these two collision systems, while the major elliptic flow-related backgrounds are expected to be the same because of the same number of nucleons~\cite{Voloshin:2010ut}. The possible nuclear deformity was found to cause only small difference in the eccentricity ($\etwo$)~\cite{Deng:2016knn}. Most of the calculations so far and the above expectations are based on the Woods-Saxon (WS) density distributions. However, WS is only an approximation to nuclear density distributions that are ultimately determined by nuclear and Coulomb interactions among the nucleons~\cite{Dreizler1990nuclear,Brown:1988vm}. The different numbers of protons and neutrons in Ru and Zr inevitably force their distributions to differ. The recent density functional theory (DFT) calculations of their distributions indicate that the collision geometries could cause sizable difference in their flow backgrounds to the CME~\cite{Xu:2017zcn}.

It is important to determine which of the nuclear density distributions--the DFT or the WS--is more trustworthy. Isobaric collisions are the best place to do so because they are highly similar except for the slight difference in the initial conditions~\cite{Xu:2017zcn}. Relative measurements of isobaric collisions are highly sensitive to those initial conditions. In this work, we expand our study in Ref.~\cite{Xu:2017zcn} using the "a multi-phase transport" ( \ampt\ ) model with the DFT nuclear densities. For comparisons, we also include \ampt\ simulations with the WS densities. The latter have recently been performed by Ref.~\cite{Deng:2018dut} focusing on the CME. In this paper we focus on standard observables. The goal is to determine, by comparing to upcoming experimental data of isobaric collisions, the correct nuclear density distributions. This shall then pave the way for further studies of isobaric collisions. In particular we contrast the multiplicity distributions and the elliptical anisotropies between Ru+Ru and Zr+Zr collisions, which can decisively determine the density distributions of the colliding nuclei. 

\section{Nuclear densities}
In relativistic heavy-ion collision studies, usually the WS nuclear density is used~\cite{Miller:2007ri,Loizides:2014vua}:
\begin{equation}
\rhows(r,\theta)=\frac{\rho_0}{1+\exp[(r-R_0[1+\beta_2Y_2^0(\theta)])/a]}\;,
\label{eq:ws}
\end{equation}
where $R_0$ is the radius parameter, $a$ is the skin diffusion thickness, $Y_2^0$ is a spherical harmonic, $\beta_2$ is the deformity quadrupole parameter, and $\rho_0$ is the normalization factor. Usually no distinction is made between proton and neutron density distributions. The charge radius parameters for Ru and Zr are listed in Table~\ref{tab:R}. These values are often taken as the mass radii used in the WS formula of Eq.~(\ref{eq:ws}) because of the lack of their experimental measurements~\cite{Deng:2016knn}; we refer to this WS version as ``\wsQ''. The thickness is taken to be $a=0.46$~fm for both Ru and Zr~\cite{Deng:2016knn}.
The $\beta_2$ values for Ru and Zr have large uncertainties. However, it was shown that even extreme cases of $\beta_2$ values introduce only small difference in $\etwo$ in mid-central to central collisions~\cite{Deng:2016knn}.
\begin{table}
  \caption{Effective nuclear radius parameters (in fm) corresponding to the WS density formula [Eq.~(\ref{eq:ws}) with $a=0.46$~fm and $\beta_2=0$].}
  \label{tab:R}
     \begin{tabular}{cc|cc|cc}\\\hline
    & & \multicolumn{2}{c|}{\Ru} & \multicolumn{2}{c}{\Zr}\\
    & & charge & mass & charge & mass \\ \hline
    \multirow{2}{*}{\wsQ} & $R_0$ & \multicolumn{2}{c|}{5.085~\cite{Deng:2016knn}} & \multicolumn{2}{c}{5.020~\cite{Deng:2016knn}} \\
    & $\sqrt{\mean{r^2}}$ & \multicolumn{2}{c|}{4.294} & \multicolumn{2}{c}{4.248} \\ \hline
    \multirow{2}{*}{DFT} & $\sqrt{\mean{r^2}}$ & 4.327 & 4.343 & 4.271 & 4.366 \\
    & $R_0\equiv1.183\sqrt{\mean{r^2}}$ & 5.119 & 5.138 & 5.053 & 5.165 \\ \hline
  \end{tabular}
\end{table}

\begin{figure}[hbt]
  \centering
  \includegraphics[width=0.5\textwidth]{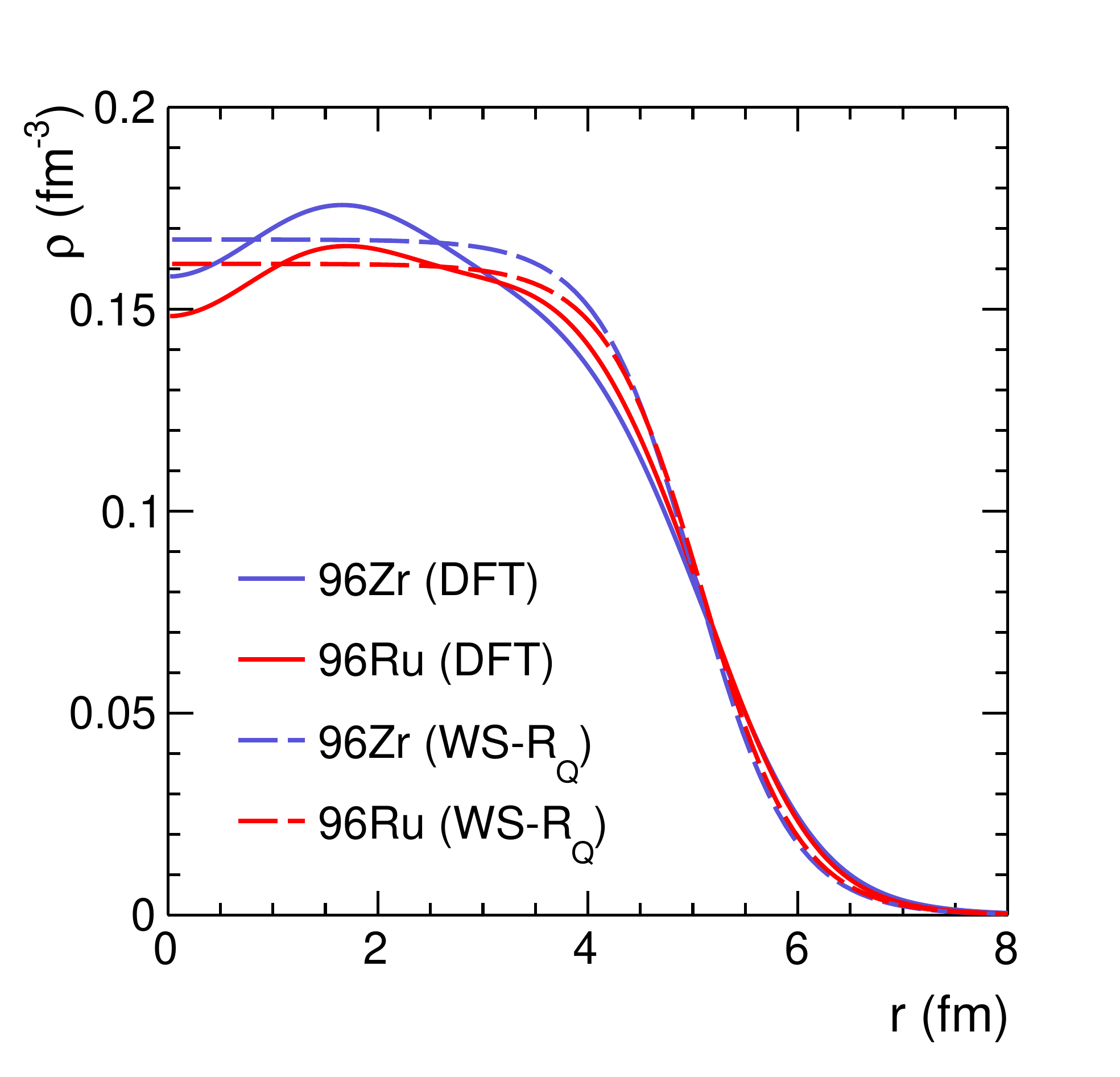}
  \vspace*{-0.3in}
  \caption{The nucleon densities of \Ru\ and \Zr\ calculated by DFT~\cite{Xu:2017zcn} (solid curves). The \wsQ\ densities~\cite{Deng:2016knn} (dashed curves) are shown for comparison.}
  \label{fig:rho}
\end{figure}

The most commonly used framework to calculate nuclear structure is the DFT~\cite{Bender:2003jk,Erler:2012xxx}.
It employs energy density functionals which incorporate complex many-body correlations that are primarily constrained by global nuclear properties such as binding energies and radii~\cite{Bender:2003jk,Erler:2012xxx,Hagen:2015yea}.
The Ru and Zr proton and neutron distributions were calculated in Ref.~\cite{Xu:2017zcn}. Figure~\ref{fig:rho} shows the nucleon densities from DFT. For comparison the \wsQ\ nucleon densities are also shown in Fig.~\ref{fig:rho}.

In order to get an intuititive comparison between the DFT and WS densities, we calculate the root-mean-square (RMS) radius, $\sqrt{\mean{r^2}}$, of the DFT proton and nucleon distributions~\cite{Xu:2017zcn}, as well as the WS distributions. The values are listed in Table~\ref{tab:R}. 
To obtain an effective $R_0$ parameter as in Eq.~(\ref{eq:ws}) for the DFT distributions, we simply multiply the $\sqrt{\mean{r^2}}$ values by the factor 1.183 (i.e.~the average ratio of $R_0/\sqrt{\mean{r^2}}$ from the \wsQ\ densities).

The DFT charge radius of Ru is larger than that of Zr because Ru has four more protons. Because of the smaller number of neutrons in Ru, its neutron radius is smaller 
than that of Zr; actually DFT results show that it is significantly smaller. As a result, the overall mass radius (i.e.~of all nucleons) of Ru is slightly smaller than that of Zr. 
On the other hand, in the WS densities~\cite{Deng:2016knn}, the charge radii are taken as the mass radii, so the Ru radius is larger than the Zr radius, which is the opposite to our DFT finding.
This is one important distinction between the DFT and \wsQ\ densities.

\section{The \ampt\ model}
We employ the string melting version of \ampt\ (\amptsm)~\cite{Lin:2001zk,Lin:2004en} in our study. The model consists of a fluctuating initial condition, parton elastic scatterings, quark coalescence for hadronization, and hadronic interactions. 
The initial condition of \ampt\ is based on the \hij\ model~\cite{Gyulassy:1994ew}, which uses the {\em Monte Carlo} Glauber (\mcg) model for the nuclei. 
We implement our DFT nuclear densities into the \hij\ component in \ampt\ (version 2.26t7). 
We refer to this as ``\amptdft.''
For comparison purposes, we also run \amptsm\ using the WS densities with the radius parameters from Table~\ref{tab:R}: One is \wsQ, and the other is the WS using the effective mass radii from the DFT densities which we refer to as ``\wsR.'' We refer to these simulations using \wsQ\ and \wsR\ as ``\amptwsQ'' and ``\amptwsR'', respectively.

The initial energy and particle productions in \ampt\ are based on \hij, where a given nuclear density distribution leads to corresponding distributions of the number of wounded (participant) nucleons ($\Npart$) and the number of binary collisions ($\Nbin$).
\amptsm\ then converts these initial hadrons into their valence quarks and antiquarks, based on the assumption that the parton
degrees of freedom are required to describe the early stage of high energy heavy-ion collisions~\cite{Lin:2001zk,Lin:2004en}.
The (anti)quarks further evolve via two-body elastic scatterings, treated with Zhang's parton cascade~\cite{Zhang:1997ej}.
The Debye-screened differential cross-section $d\sigma/dt\propto\alpha_s^2/(t-\mu_D^2)^2$~\cite{Lin:2004en} is used in \ampt, with the strong coupling constant $\alpha_s=0.33$ and Debye screening mass $\mu_D=2.265$/fm resulting in a total parton scattering cross section of $\sigma=3$~mb. After partons stop interacting, a simple quark coalescence model is applied to convert partons into hadrons~\cite{Lin:2004en}. Subsequent interactions of those formed hadrons are modeled by a hadron cascade~\cite{Lin:2004en}. We terminate the hadronic interactions at a cutoff time of 30~\fmc. 

Hadronization in \amptsm\ is modeled with a simple quark coalescence, where two nearest partons in coordinate space (one quark and one antiquark) are combined into a meson and three nearest quarks (or antiquarks) are combined into a baryon (or antibaryon). In addition, when the flavor composition of the coalescing quark and antiquark allows the formation of either a pseudo-scalar or a vector meson, the meson species whose mass is closer to the invariant mass of the coalescing parton pair will be formed. The same criterion is also applied to the formation of an octet or a decuplet baryon with the same flavor composition. 
The hadron cascade in \ampt\ includes explicit particles such as $\pi$, $\rho$, $\omega$, $\eta$, $K$, $K^*$, $\phi$ mesons, $N$, $\Delta$, $N^*(1440)$, $N^*(1535)$, $\Lambda$, $\Sigma$, $\Xi$, $\Omega$ baryons and antibaryons, plus deuterons and antideuterons~\cite{Oh:2009gx}. 
Hadronic interactions include meson-meson, meson-baryon, and baryon-baryon elastic and inelastic scatterings. More details can be found in Ref.~\cite{Lin:2004en}.

\section{Model predictions}
We simulate a total of 80, 35, and 32 million minimum-bias (MB) events each for Ru+Ru and Zr+Zr collisions using \amptdft, \amptwsQ, and \amptwsR, respectively. The impact parameter ($b$) range is set to be 0--20~fm.

\begin{figure}[hbt]
  \centering
  \includegraphics[width=0.5\textwidth]{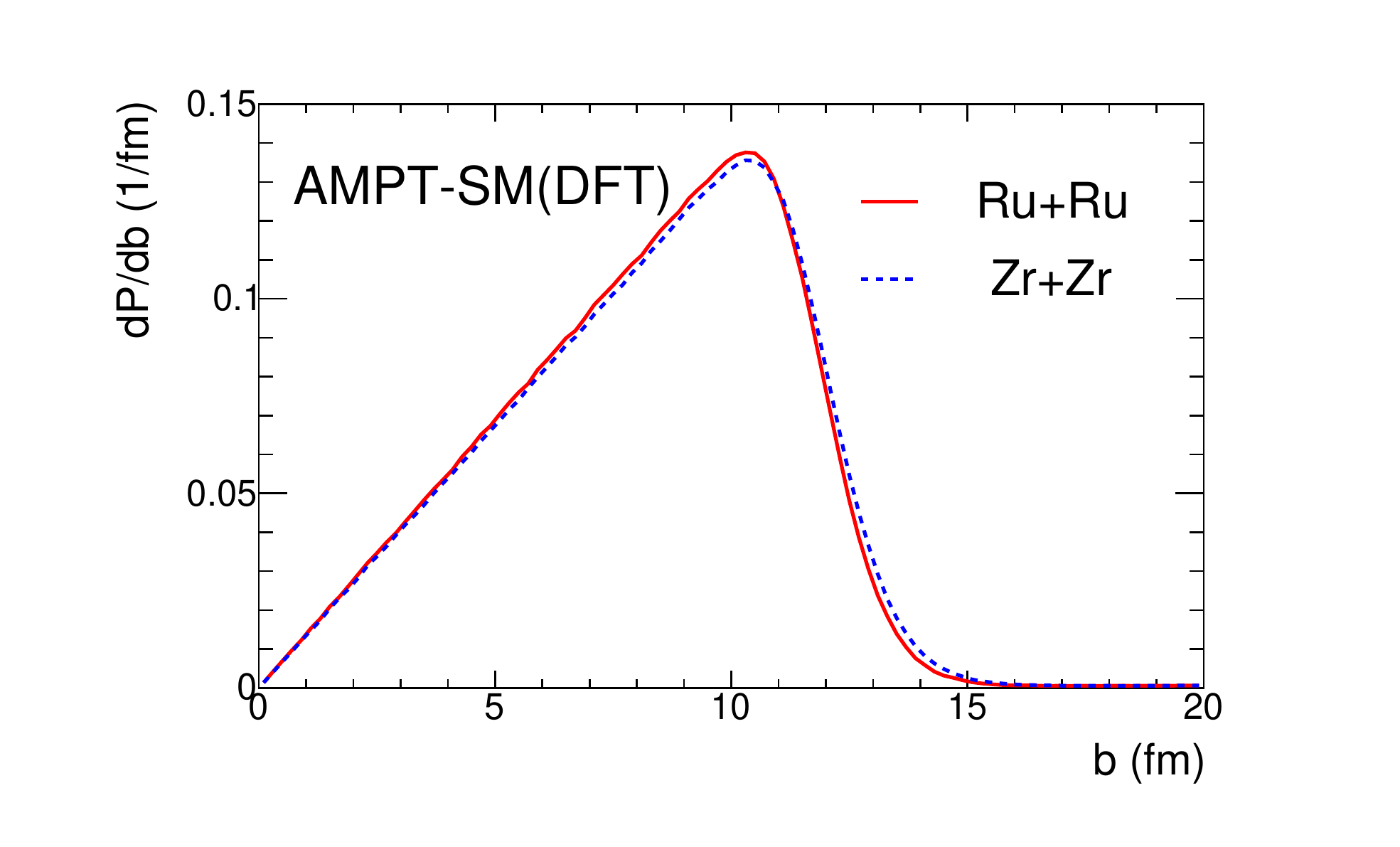}
  \vspace*{-0.3in}
  \caption{Impact parameter ($b$) probability distributions in Ru+Ru and Zr+Zr collisions simulated by \amptdft.
  Proton and neutron density distributions of the \Ru\ and \Zr\ nuclei are assumed spherical and calculated by DFT.}
  \label{fig:b}
\end{figure}

Figure~\ref{fig:b} shows the probability distributions in $b$ from \amptdft\ for an interaction to happen ($\Npart\geq2$). The probability is linear in $b$ up to 10~fm or so because at small $b$ the two colliding nuclei are guanranteed to interact. At larger $b$ the probability drops because not every encounter can have at least one nucleon-nucleon (NN) interaction. The drop happens at slightly smaller $b$ in Ru+Ru than Zr+Zr collisions because Ru is slightly smaller than Zr from the DFT calculations.

\subsection{Multiplicity distributions}\label{sec:mult}
\begin{figure}[hbt]
  \centering
  \includegraphics[width=0.5\textwidth]{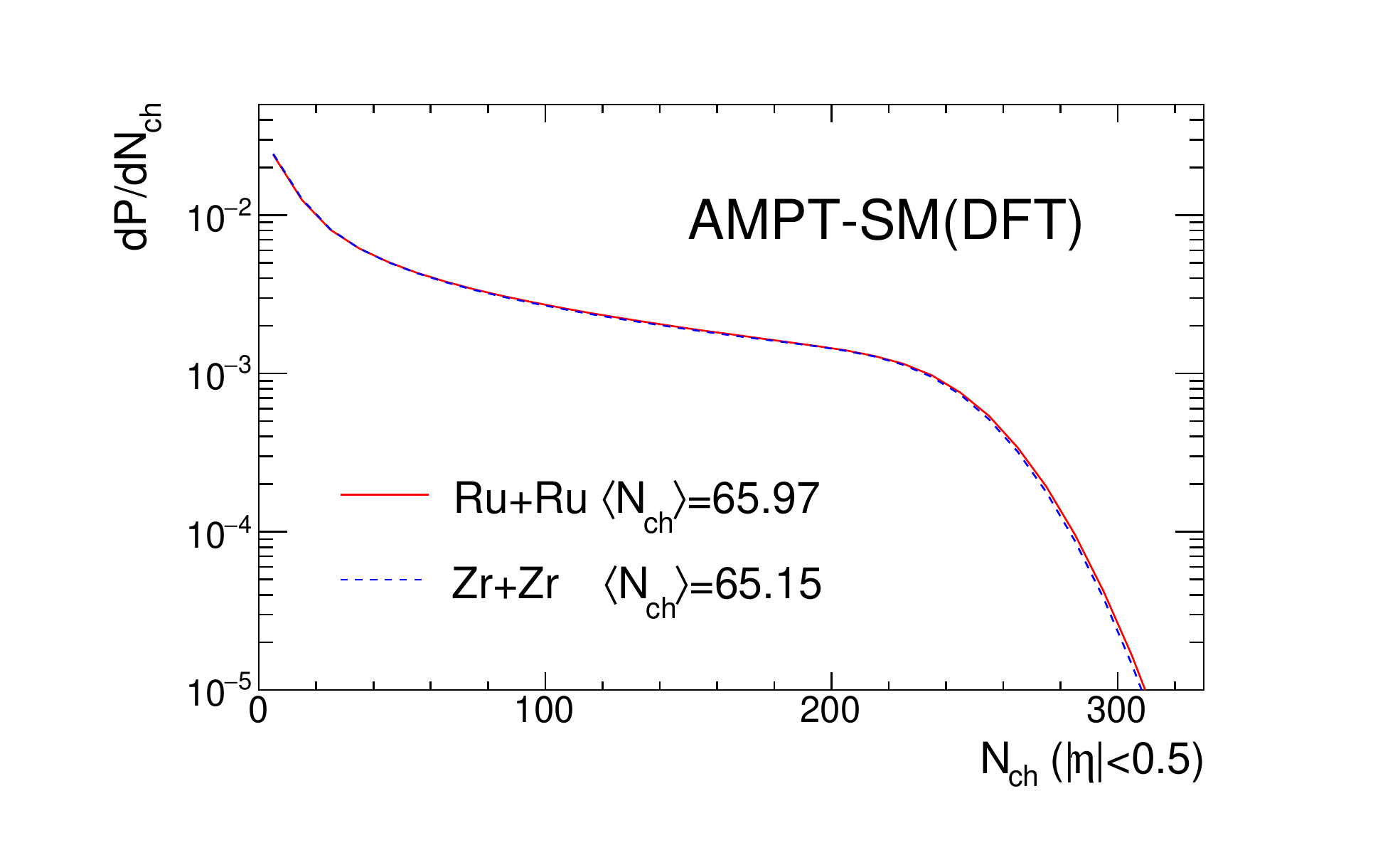}
  \vspace*{-0.3in}
  \caption{Midrapidity charged hadron multiplicity ($\Nch$) distributions in Ru+Ru and Zr+Zr collisions simulated by \amptdft. $\Nch$ is the sum of charged pion, charged kaon, proton and antiproton multiplicities within $|\eta|<0.5$ and $\pt > 0.2 $ \gevc.}
  \label{fig:Nch}
\end{figure}

Figure~\ref{fig:Nch} shows the midrapidity charged hadron multiplicity ($\Nch$) distributions from the \amptdft\ simulations. The multiplicity counts all charged pions, charged kaons, protons and antiprotons within the pseudorapidity range of $|\eta|<0.5$ and transverse momentum range of $\pt > 0.2 $ \gevc.
The distributions are almost identical between Ru+Ru and Zr+Zr collisions. The average multiplicities $\mean{\Nch}$ for MB Ru+Ru and Zr+Zr collisions are 65.97 and 65.15, respectively. The slightly larger $\mean{\Nch}$ and larger $\Nch$ tail in Ru+Ru than Zr+Zr collisions is because of the smaller effective radius of Ru than Zr from the DFT mass densities~\cite{Xu:2017zcn}. With smaller radius, the Ru+Ru collision zone is smaller to start with, and the energy density should be higher and the number of binary collisions larger.

\begin{figure}[hbt]
  \centering
  \includegraphics[width=0.54\textwidth]{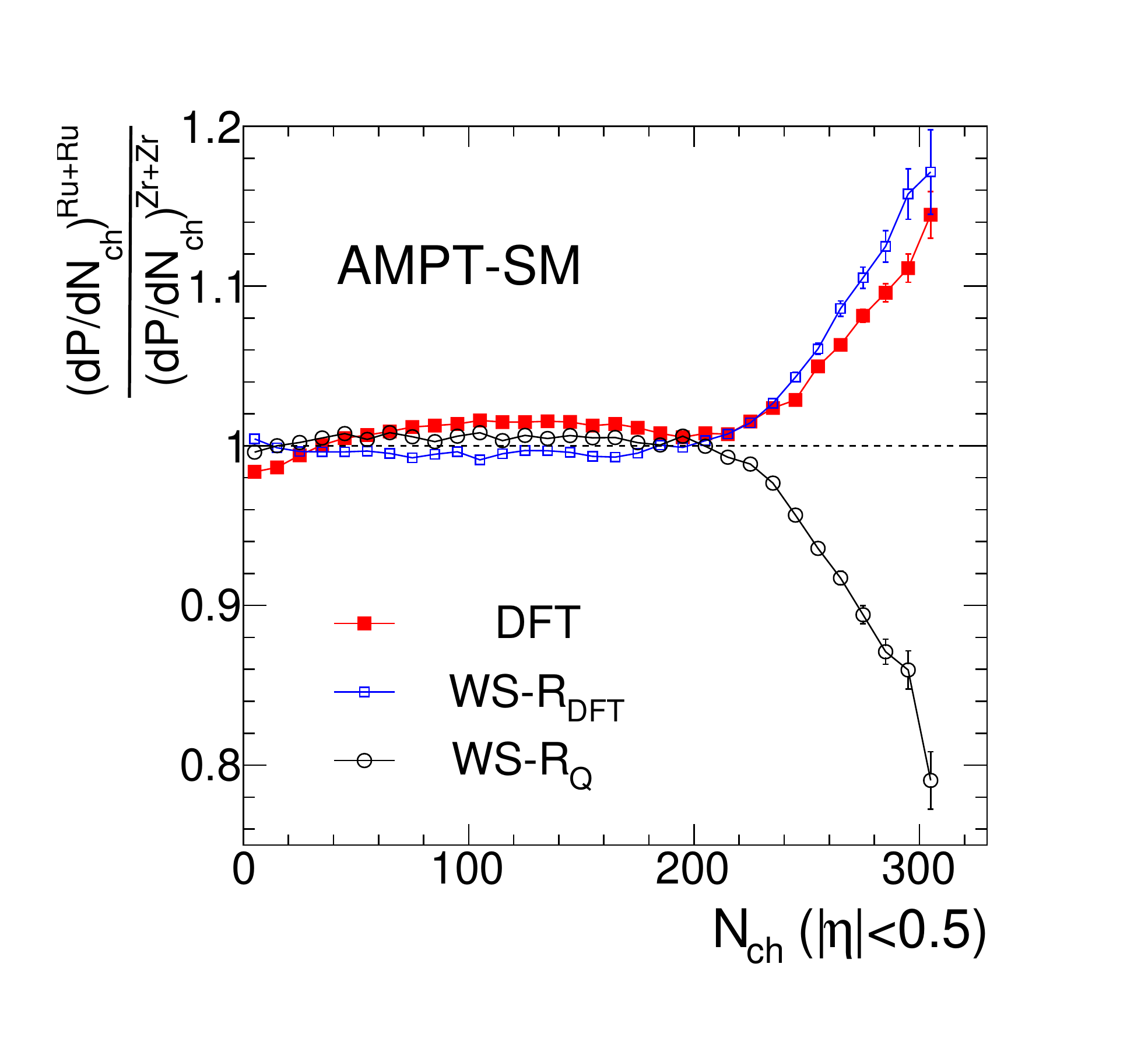}
  \vspace*{-0.3in}
  \caption{Ratio of the $\Nch$ distributions in Ru+Ru and Zr+Zr collisions simulated by \amptsm. Three types of nuclear density distributions are shown: DFT~\cite{Xu:2017zcn}, \wsQ, and \wsR.}
  \label{fig:Nratio}
\end{figure}

Figure~\ref{fig:Nratio} shows the ratio of the $\Nch$ distribution in Ru+Ru to that in Zr+Zr collisions. The DFT density result is shown in red where the ratio curves up at large $\Nch$ because of the larger $\Nch$ tail in Ru+Ru (see Fig.~\ref{fig:Nch}). Also shown for comparison is the ratio from the \wsQ\ densities, which is consistent with the calculation in Ref.~\cite{Deng:2016knn}. The trend of the ratio in the \wsQ\ case is the opposite to the DFT case, because the Ru charge radius is larger than the Zr's in the \wsQ.
These opposite behaviors are a decisive discriminator for the relative mass radii between the Ru and Zr nuclei.
Also shown in Fig.~\ref{fig:Nratio} is the ratio from simulations using the \wsR, which give a similar trend as \amptdft\ that directly used DFT densities. This confirms the conclusion that the tail behavior in the ratio is mainly due to the ordering of the nuclear mass radii.

There is observable difference in the intermediate $\Nch$ region between simulations directly using DFT densities and using WS densities with DFT effective radii. This may be used to tell whether the density distribution is DFT like or simply WS like with the correct ordering of the mass radii. In fact, the intermediate $\Nch$ regions of the ratios from the two WS densities appear unfeatured, however, that from the DFT density seems nontrivially structured. If experimentally confirmed, then it would constitute strong evidence for the validity of the DFT densities for the isobaric nuclei. 

A very successful isobar run has just concluded in 2018. STAR has taken approximately 2 billion minimum bias events each for Ru+Ru and Zr+Zr collisions. This is a factor of 25 higher statistics compared to the AMPT statistics we have run for AMPT-SM(DFT). The data statistical error bars would be approximately a factor of 5 smaller than those shown in Fig.~\ref{fig:Nratio}. Given the isobar beam qualities and the frequent alternating of the Ru and Zr beams during the run, the experimental systematic uncertainties are expected to be small in the ratio of the multiplicity distributions. With the large, opposite behaviors of the ratios at large $\Nch$ between the DFT and WS densities, there should be no ambiguity to distinguish them. Even the intermediate $\Nch$ range may have enough discrimination power.


\subsection{Centrality definition and Glauber calculations}\label{sec:Glauber}
We define centrality using midrapidity ($|\eta|<0.5$) charged hadron multiplicity in $\pt > 0.2 $ \gevc\ as shown in Fig.~\ref{fig:Nch}, similarly to experimental data analysis~\cite{Abelev:2008ab}. We determine the multiplicity ranges of 20 centrality bins of 5\% equal size. The most central bin is referred to as 0-5\% (or top 5\%). Because of the slight difference in the multiplicity distributions, the multiplicity ranges for a given centrality bin can be slightly different between Ru+Ru and Zr+Zr collisions.

With the distributions of impact parameters in each centrality multiplicity bin obtained from the \amptsm\ model, we use a \mcg\ model~\cite{Alver:2006wh,Miller:2007ri,Rybczynski:2011wv,Xu:2014ada,Zhu:2016puf} to simulate the initial geometry of each isobaric collision in a given centrality bin. In our \mcg\ model, the locations of protons and neutrons in a nucleus are generated according to the given (DFT or WS) proton and neutron density profiles. The minimum internucleon distance is set to be $d_{\rm min} = 0.4$ fm and the nucleon-nucleon cross section is set to be  $\sigma_{\rm NN} = 42$ mb. Instead of the hard-sphere approximation with a traditional step function,  a participant nucleon is determined by a Gaussian-type nucleon-nucleon collision profile $p(b)= A\exp{(-\pi Ad^2/\sigma_{\rm NN})}$, where $d$ is the relative transverse distance between the nucleon and the surrounding nucleons from the other nucleus and $A=0.92$~\cite{Rybczynski:2011wv}. We then calculate the average number of participant nucleons ($\Npart$) and the average number of binary nucleon-nucleon collision ($\Nbin$) for each centrality bin. 

The initial geometric anisotropy of the transverse overlap region of a heavy-ion collision is often described by eccentricity of the $n$th-harmonic order~\cite{Alver:2010gr}:
\begin{equation}
  \epsilon_n=\left.\sqrt{\mean{\rt^2\cos n\phi_r}^2+\mean{\rt^2\sin n\phi_r}^2}\right/\mean{\rt^2}\,.
\end{equation}
Here $\rt$ and $\phi_r$ are the polar coordinate of each participant nucleon in the transverse plane, and $\mean{...}$ denotes the per-event average.

\begin{figure}[hbt]
  \centering
  \includegraphics[width=0.45\textwidth]{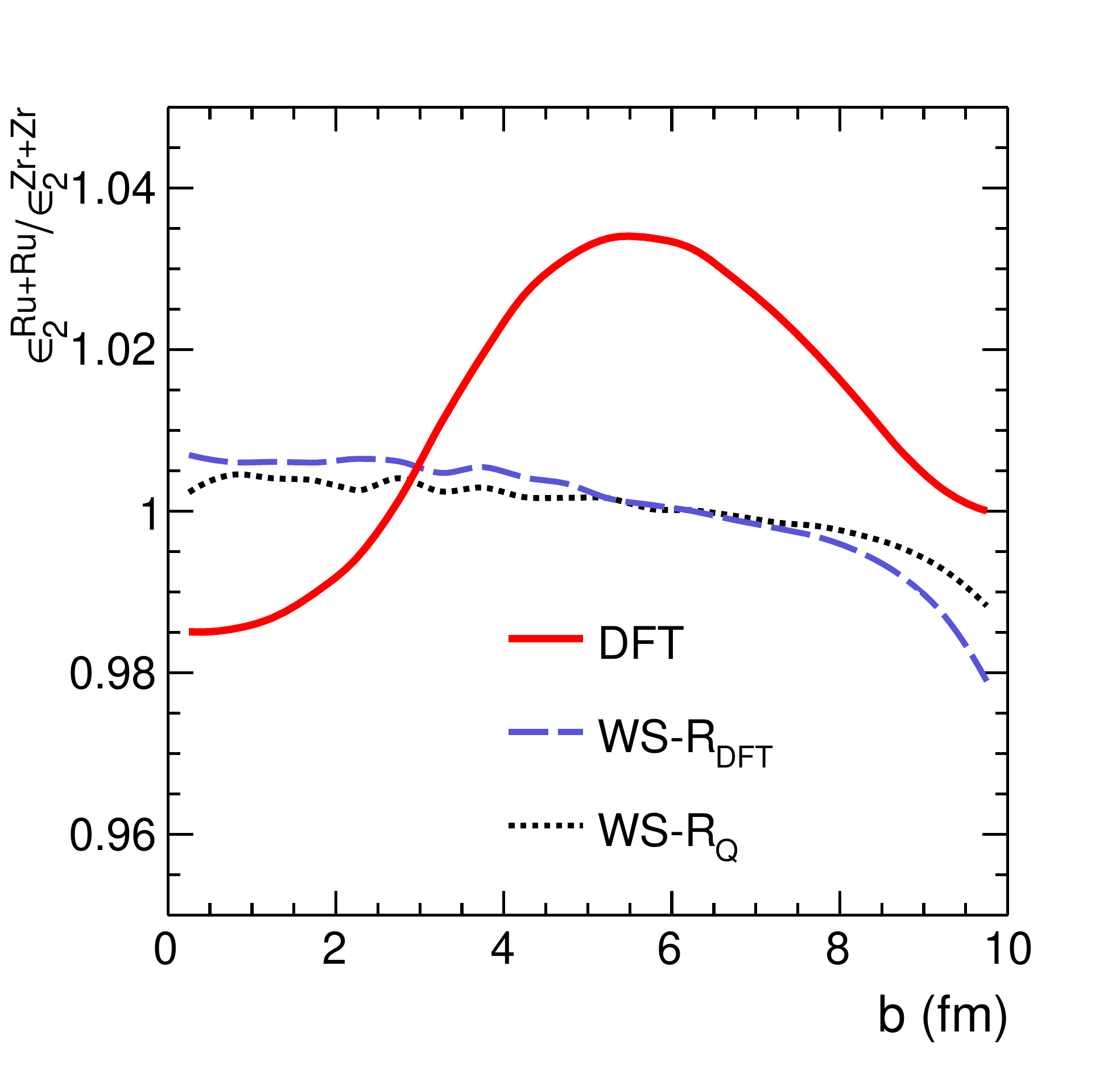}
  \caption{Eccentricity ratios in Ru+Ru to Zr+Zr collisions calculated by \mcg. Three types of nuclear density distributions are shown: DFT~\cite{Xu:2017zcn}, \wsQ\ (Table~\ref{tab:R}), and \wsR\ (Table~\ref{tab:R}).}
  \label{fig:en}
\end{figure}

Figure~\ref{fig:en} shows the ratio of eccentricities in Ru+Ru to Zr+Zr collisions as a function of $b$. Three types of the density distributions are displayed. The two WS density distributions give similar ratios. The DFT density gives quite different eccentricities for Ru+Ru and Zr+Zr; the ratio significantly deviates from unity~\cite{Xu:2017zcn}. This indicates that the eccentricity is sensitive to the internal structure of the colliding nuclei. It can be used to tell apart the DFT and WS density scenarios with the same effective radii, where the multiplicity distributions may lose distinguishing power (see Sect.~\ref{sec:mult}). The experimental quantity to tell them apart is the elliptical anisotropy $v_2$ (as $v_2$ is proportional to $\etwo$, discussed in Sect.~\ref{sec:vn}). 

\subsection{Particle production}\label{sec:prod}
\begin{figure}[hbt]
  \centering
  \includegraphics[width=0.5\textwidth]{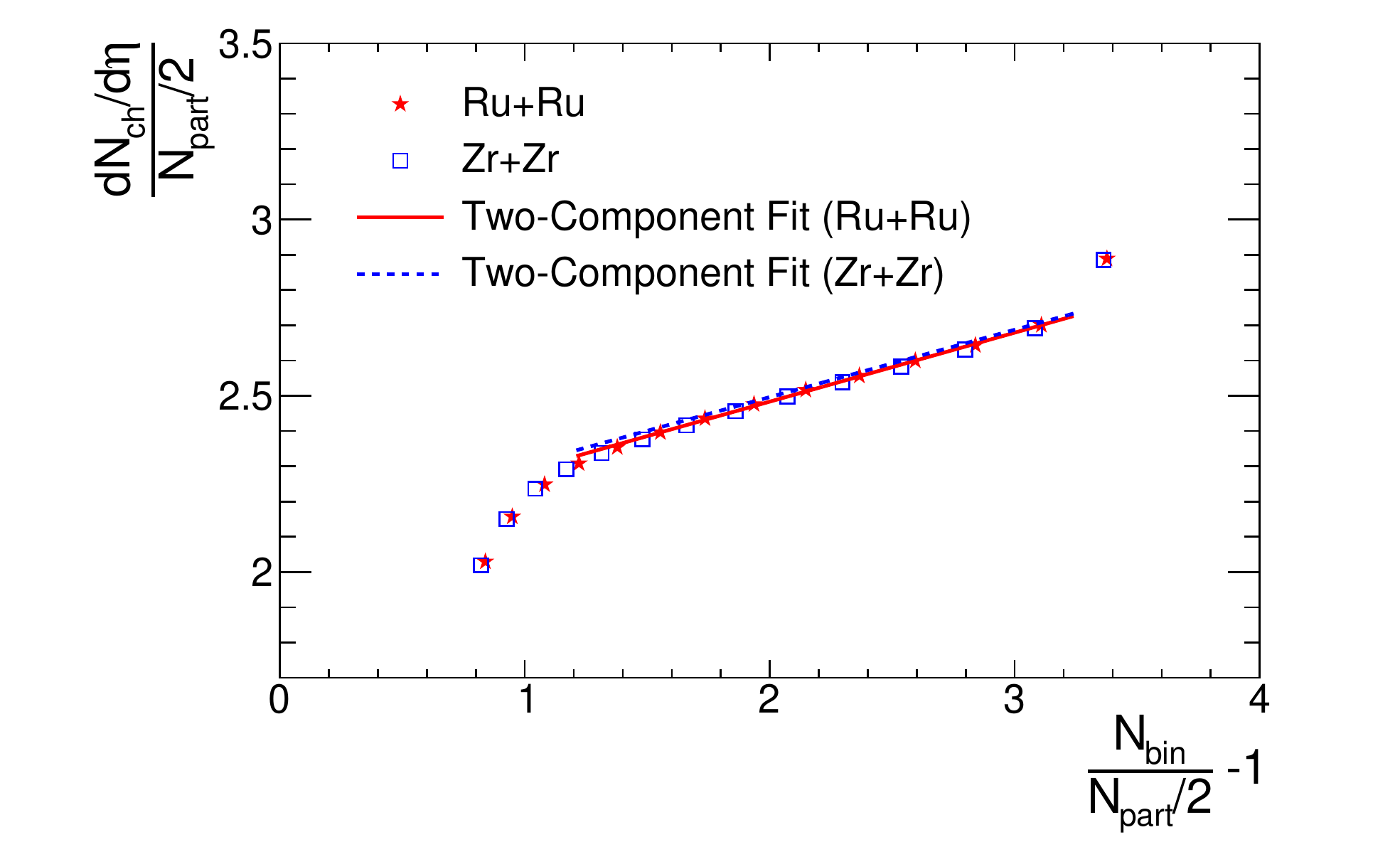}
  \caption{Charged particle multiplicity per participant pair as a function of $\frac{\Nbin}{\Npart/2}-1$ in Ru+Ru and Zr+Zr collisions simulated by \amptdft. The star and open square represent AMPT results. The line are two-component model fits to the 5-50\% centrality range.}
  \label{fig:2comp_fit}
\end{figure}

Figure~\ref{fig:2comp_fit} shows the charged hadron multiplicity per participant pair as a function of $\frac{\Nbin}{\Npart/2}-1$ in the 0-70\% centrality range where $\Npart$ and $\Nbin$ are determined by \mcg\ in Sect.~\ref{sec:Glauber}. Since the centrality is defined by multiplicity, there is always a multiplicity bias in peripheral and central collisions. The bias is evident for the top 5\% centrality where the data point does not follow the trend of the other data points. The trend bends down in peripheral collisions, likely also because of multiplicity biases. We defer detailed studies of multiplicity biases in centrality definitions to a future work.

We fit the 5-50\% centrality range where multiplicity biases are minimal by a two-component model as in Ref~\cite{Kharzeev:2000ph,Wang:2000bf,Back:2004dy}: 
\begin{equation}
    \frac{\dNdeta}{\Npart/2}=n_{pp}\left[1+x\left(\frac{\Nbin}{\Npart/2}-1\right)\right]\,.
  \end{equation}
The $n_{pp}$ is the corresponding charge multiplicity in $p+p$ collisions and the fraction $x$ represents the contribution to the multiplicity from ``hard processes''. The fitting results are shown in Fig.~\ref{fig:2comp_fit}. The fit parameters are $n_{pp}$ $\approx$ 2.1 and $x \approx 9\%$ for both Ru+Ru and Zr+Zr. These values are in the ballpark of those extracted experimentally at RHIC~\cite{Back:2004dy}.
  
\begin{figure}[hbt]
  \centering
  \includegraphics[width=0.5\textwidth]{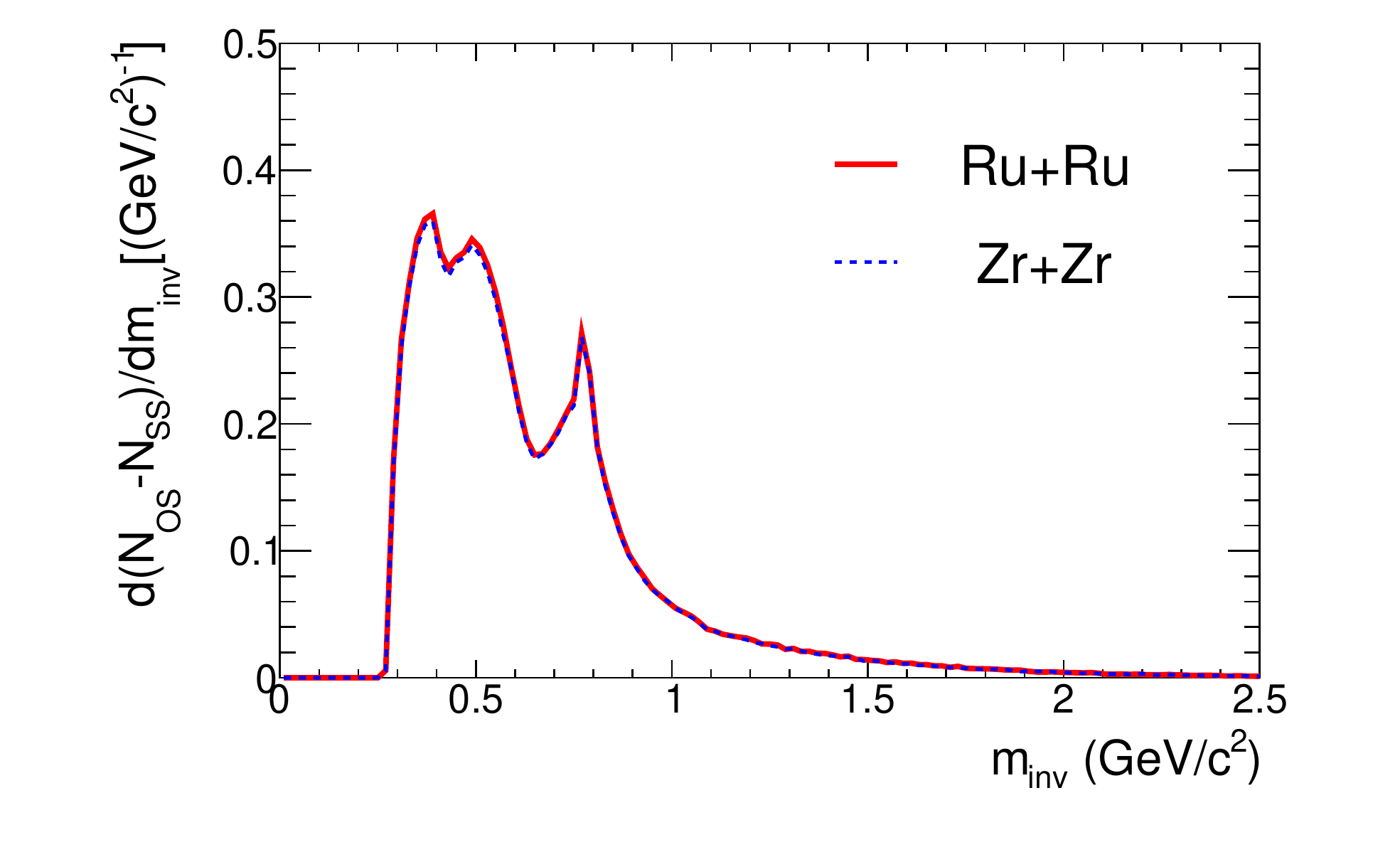}
  \caption{Difference of the invariant mass ($\minv$) distributions of opposite-sign (OS) pion pairs and same-sign (SS) pion pairs in MB Ru+Ru and Zr+Zr collisions simulated by \amptdft. Pions within $|\eta|<1$ are used.}
  \label{fig:minv}
\end{figure}

We have presented predictions on the production of common charged hadrons (mainly pions, kaons, protons and antiprotons). It is interesting to also examine short lived resonances. This is especially interesting for the isobaric collisions because resonance decays present major backgrounds to the CME~\cite{Zhao:2018ixy,Wang:2009kd,Pratt:2010zn,Bzdak:2010fd,Wang:2016iov, Zhao:2018skm,Zhao:2017nfq,Li:2018oot}. Figure~\ref{fig:minv} shows the difference of the opposite-sign and same-sign pion pair invariant mass (minv) distributions in MB Ru+Ru and Zr+Zr collisions. 
The $\minv$ distributions are nearly identical and the ratio is flat. The average ratio of the distribution in Ru+Ru to that in Zr+Zr is approximately $1.013\pm0.007$, consistent with the average $\mean{\Nch}$ ratio (see Fig.~\ref{fig:Nch}). Note that the shape of Fig.~\ref{fig:minv} is somewhat different from Ref.~\cite{Li:2018oot} where the hadronic rescatterings were not included.

\subsection{Elliptic anisotropy}\label{sec:vn}
There are many methods to analyze anisotropic flows ($v_n$) in heavy ion collisions~\cite{Poskanzer:1998yz}. We use the event plane (EP) method in this study as commonly used in experiments. Our main objective is to investigate the relative $v_2$ magnitudes between Ru+Ru and Zr+Zr collisions. The relative $v_2$ magnitudes are insensitive to analysis methods.

We reconstruct the event plane (EP) from the final-state particle momentum distribution~\cite{Poskanzer:1998yz}:
\begin{equation}
\psi_2=\frac{1}{2}{\rm atan2}(\mean{\sin 2\phi},\mean{\cos 2\phi})\,,
\end{equation}
where $\phi$ is the particle azimuthal angle,
and atan2($\mean{\sin 2\phi},\mean{\cos 2\phi}$) returns the four-quadrant inverse tangent of $\mean{\sin 2\phi}/\mean{\cos 2\phi}$. Due to finite multiplicity, the reconstructed EP is not 100\% precise. An correction is applied to $v_2$ for the EP resolution $R_2$:
\begin{equation}
  v_2=\mean{\cos2(\phi-\psi_2)}/R_2\,.
\end{equation}
The resolution is obtained by an iterative procedure using the subevent method~\cite{Poskanzer:1998yz}.

\begin{figure}[hbt]
  \centering
  \includegraphics[width=0.45\textwidth]{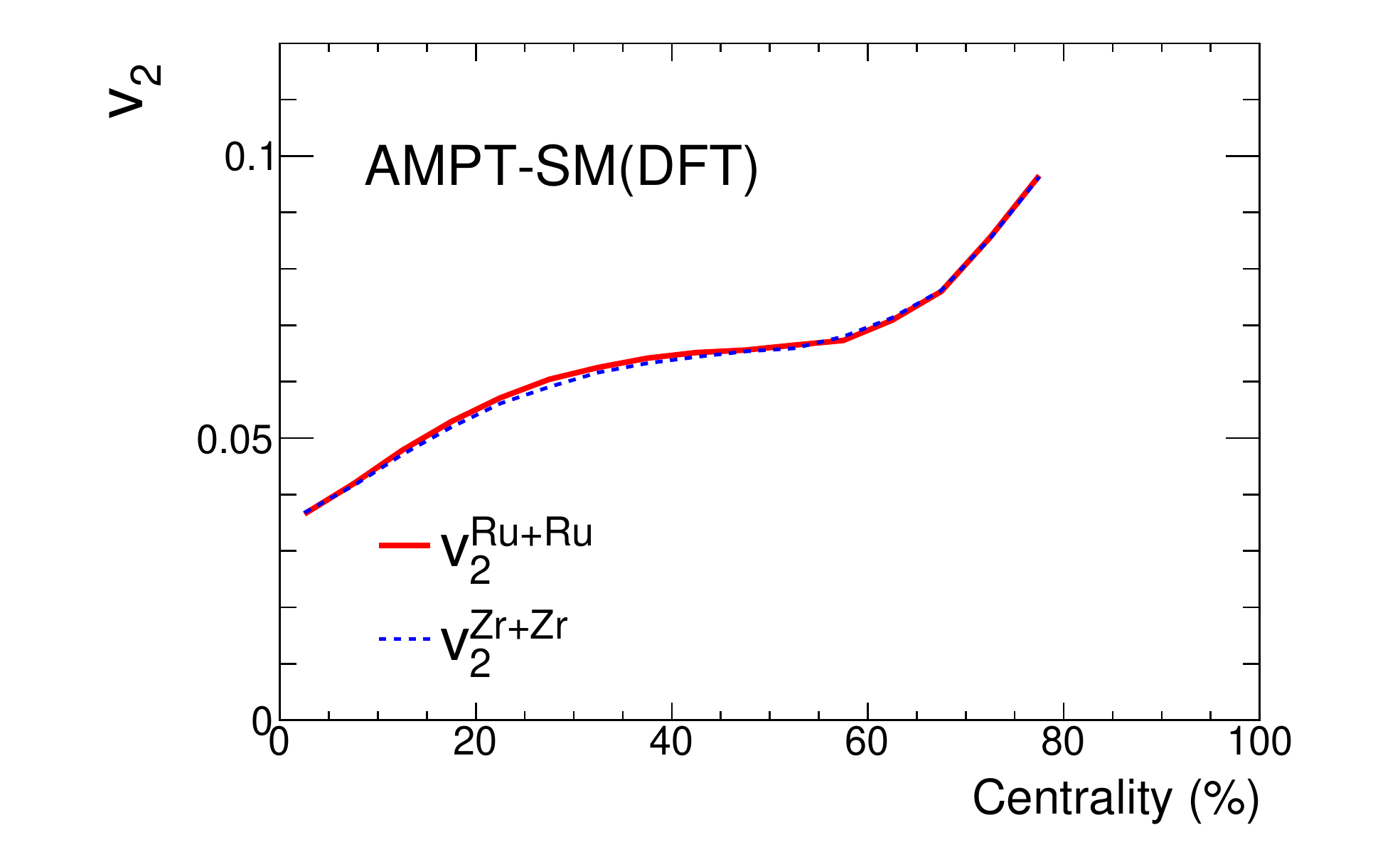}
  \caption{Azimuthal anisotropies $v_2$ of charged hadrons (within $|\eta|<1$) with respect to the EP in Ru+Ru and Zr+Zr collisions as a function of centrality, simulated by \amptdft.}
  \label{fig:vnEP}
\end{figure}

The $v_2$ from the EP method is shown in Fig.~\ref{fig:vnEP} as a function of centrality. In our EP calculation, we used all particles within $|\eta|<1$, except for the particle of interest (POI) for $v_2$ calculation. There is no $\eta$ gap between the POI and the EP to suppress nonflow contributions~\cite{Abdelwahab:2014sge}, noting that nonflow is not a major effect in central collisions because of the large event multiplicity. It is, however, a significant contributor to $v_2$ in peripheral collisions.

\begin{figure}[hbt]
  \centering
  \includegraphics[width=0.5\textwidth]{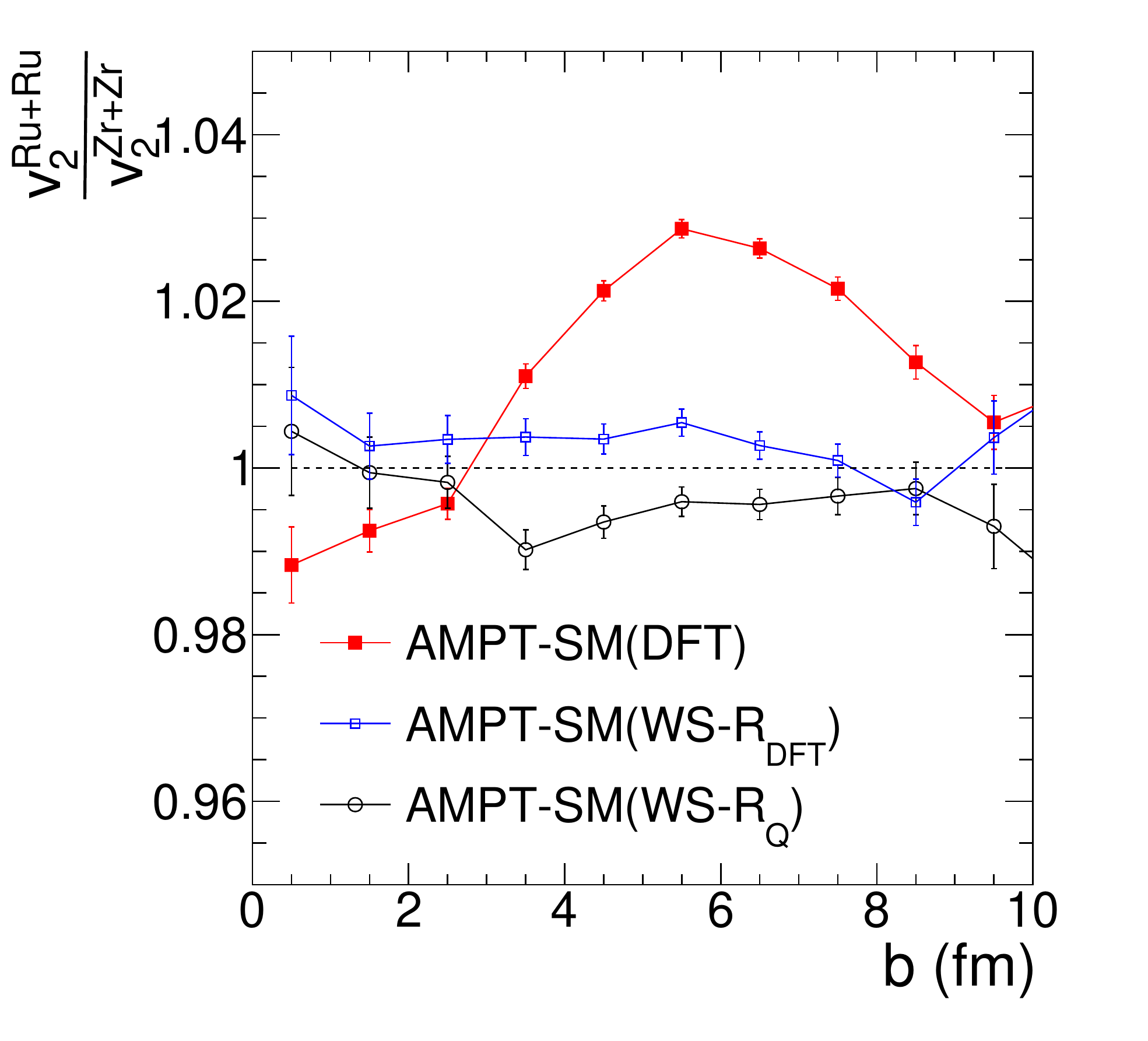}
  \caption{Ratios of charged hadron $v_2$ in Ru+Ru to Zr+Zr collisions as a function of $b$. Three nuclear density distributions are simulated: DFT, \wsQ, and \wsR.}
  \label{fig:vnratio}  
\end{figure}

As shown in Sect.~\ref{sec:Glauber}, the DFT and WS density distributions give appreciable difference in the eccentricity. This difference should be reflected in the final-state $v_2$. We show in Fig.~\ref{fig:vnratio} the $v_2$ ratios in Ru+Ru to Zr+Zr collisions for three types of density distributions as a function of $b$. Indeed, the eccentricity difference shows up in the final-state $v_2$. 

\begin{figure}[hbt]
  \centering
  \includegraphics[width=0.5\textwidth]{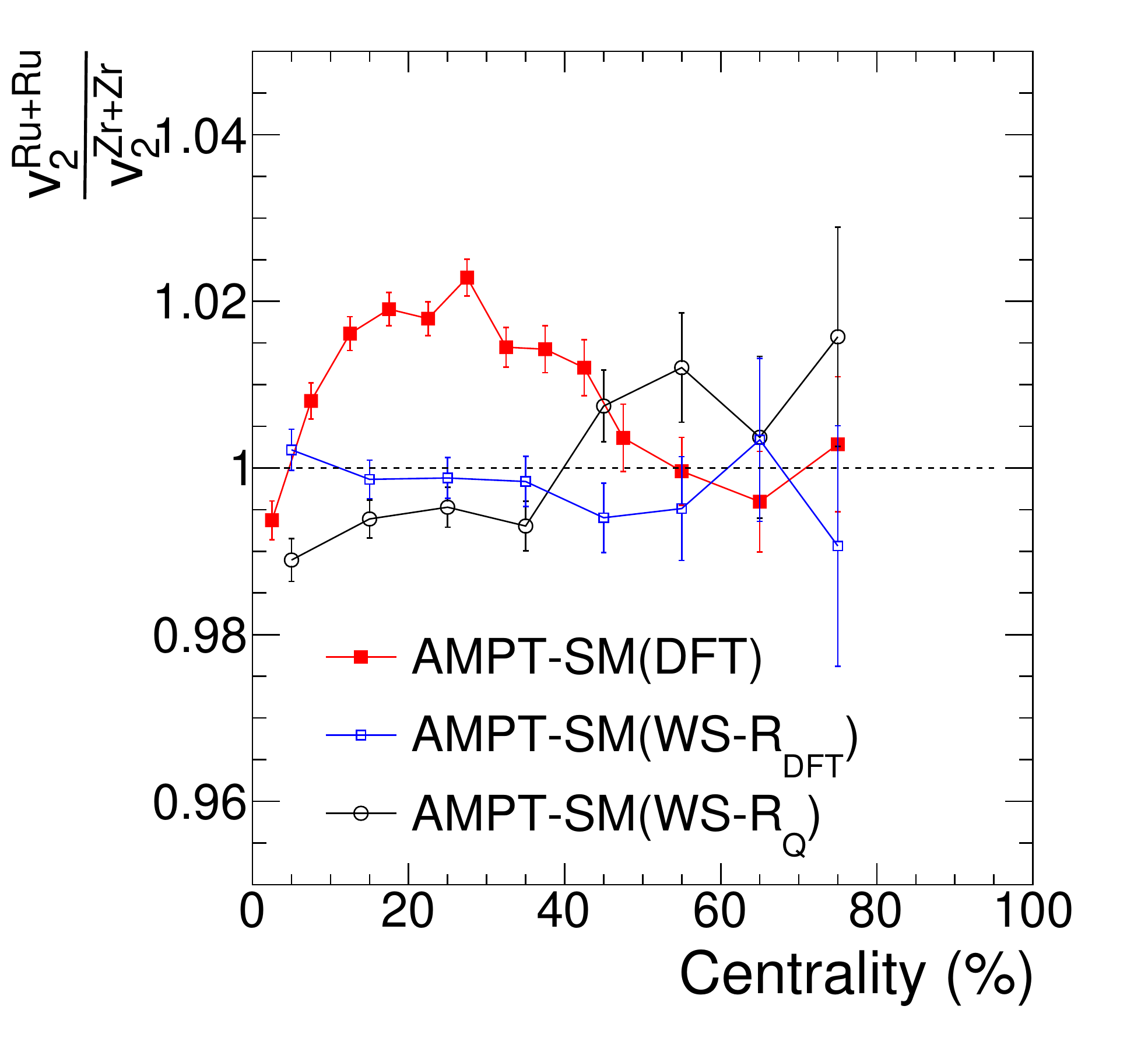}
  \caption{Ratios of charged hadron $v_2$ in Ru+Ru to Zr+Zr collisions as a function of centrality. Three nuclear density distributions are simulated: DFT, \wsQ, and \wsR.}
   \label{fig:v2ratio}
\end{figure}

Since experimentally the $v_n$ is measured against centrality, Fig.~\ref{fig:v2ratio} shows the $v_2$ ratio in Ru+Ru to Zr+Zr collisions for the three types of density distributions as a function of centrality. 
The centralites for \amptwsQ\ and \amptwsR\ are obtained in a similar way using their respective multiplicity distributions. The ratio obtained from the DFT densities is clearly different from those using the WS densities.
This can be exploited to discriminate between DFT and WS nuclear densities by comparing to the upcoming $v_2$ measurements from the isobaric collision data. 
With the approximate 25-fold higher data statistics compared to our AMPT simulation, and with largely canceled systematic uncertainties in the $v_2$ ratio, the two scenarios of DFT and WS nuclear densities can be decisively determined. Note that the difference in the DFT $v_2$ ratios between Fig.~\ref{fig:vnratio} and Fig.~\ref{fig:v2ratio} is because of the smearing in $b$ when the centrality is defined according to multiplicity.

\section{Summary}
In this paper, we make predictions of isobaric \Ru+\Ru\ and \Zr+\Zr\ collisions using the string-melting version of the \ampt\ model with the nuclear density distributions calculated by the DFT. We present the charged hadron multiplicity ($\Nch$) distributions
and the elliptic azimuthal anisotropies $v_2$ as a function of centrality. Emphases are put on the relative differences between the two isobaric systems.

We show that, while the charge radius of \Ru\ is larger than that of \Zr, the mass radius of \Ru\ from DFT is slightly smaller than that of \Zr. Because of this, the ratio of the $\Nch$ distributions in Ru+Ru to Zr+Zr collisions curves up at large $\Nch$, opposite to the trend obtained using the common WS densities with charge radii. This feature can be checked against isobaric data to decisively determine the relative mass radii of the isobaric nuclei. 

With the same effective mass radii, while the multiplicity distribution ratio may lose the discriminating power between DFT and WS densities, the centrality dependence of the $v_2$ ratio in Ru+Ru to Zr+Zr collisions can decisively determine whether DFT density is more realistic than WS.

For the CME search, it is important to first determine the initial conditions of the isobaric collisions. The importance of our work lies in the testable predictions from several viable nuclear density distributions by a commonly used transport model. With the large data statistics (billions of events) accumulated in the isobar run at RHIC and cancellation of systematic uncertainties, our predictions should be able to decisively determine the initial conditions of the isobaric collisions and pave the way for further studies, particularly in terms of the CME search.

\section*{Acknowledgments}
This work was supported in part by
the National Natural Science Foundation of China (Grants No.~11747312, No.~11790325, No.~11790323, No.~U1732138, No.~11505056, No.~11605054, and No.~11628508) and the U.S.~Department of Energy (Grant No.~de-sc0012910).

\bibliography{ref}
\end{document}